# Single Crystal Growth of Magnesium Diboride by Using Low Melting Point Alloy Flux

Wei Du[*], Xugang Xian, Xiangjin Zhao, Li Liu, Zhuo Wang, Rengen Xu

School of Environment and Material Engineering, Yantai University, Yantai 264005, P. R. China

**Abstract**

Magnesium diboride crystals were grown with low melting alloy flux. The sample with superconductivity at 38.3 K was prepared in the zinc-magnesium flux and another sample with superconductivity at 34.11 K was prepared in the cadmium-magnesium flux. $MgB_4$ was found in both samples, and $MgB_4$'s magnetization curve being above zero line had not superconductivity. Magnesium diboride prepared by these two methods is all flaky and irregular shape and low quality of single crystal. However, these two metals should be the optional flux for magnesium diboride crystal growth. The study of single crystal growth is very helpful for the future applications.

Keywords: Crystal morphology; Low melting point alloy flux; Magnesium diboride; Superconducting materials

Corresponding author. Tel: +86-13235351100

Fax: +86-535-6706038

Email-address: duwei@ytu.edu.cn **(W. Du)**

## 1. Introduction

Since the discovery of superconductivity in Magnesium diboride ($MgB_2$) with a transition temperature ($T_c$) of 39K [1], a great concern has been given to this material structure, just as high-temperature copper oxide can be doped by elements to improve its superconducting temperature. Up to now, it is failure to improve $T_c$ of doped $MgB_2$ through C [2-3], Al [4], Li [5], Ir [6], Ag [7], Ti [8], Ti, Zr and Hf [9], and Be [10] on the contrary, the $T_c$ is reduced or even disappeared.

During the study of element doped $MgB_2$, it was also trying to achieve $MgB_2$ crystal growth by using flux as the structural stability of $MgB_2$. Cho et al [11] obtained hexagonal plate shape crystal about 100$\mu$m in the presence of Na by using high pressure crystal growth technique. Souptel et al [12] synthesized a mixture of $MgB_2$ by using Mg-Cu-B flux, and $MgB_2$ crystal could not be separated. Up to now, single-phase crystals synthesis by using low melting point alloy flux under ambient pressure had not been reported. At the same time, the M-Mg-B (M for the low melting point metal) phase diagram had also not been reported. So, it is helpful to grow $MgB_2$ single crystal by using low melting point alloy flux for the new superconducting material exploration and new single crystal growth technique.

In this paper, we investigated the single crystal growth of $MgB_2$ by using low melting point alloy flux. The selection principle of alloy flux, there are two points to note: First, structure of the metal itself with the structure of $MgB_2$ as close as possible to ensure that there is the possibility of dissolution; Second, lower melting point metals, while ensuring the final product and the flux can be separated. Based on these two points, we selected the hexagonal structure of Zn, Cd and Mg Metal to form low melting point alloy as flux.

As we all know, Mg, Zn and Cd vapor pressure increases with the temperature over its melting point. All the metals presented single atomic or molecular ideal crystal properties after gasification, which satisfied physical relation between vapor pressure and temperature (only for metals):

$$\lg P^* = \frac{A}{T} + B = \frac{-\Delta_{vap}H^\theta}{2.303RT} + \frac{C}{2.303R}$$

If every thermal data of Mg, Zn and Cd are provided for the above-mentioned equation, we can conclude that $MgB_2$ crystals grow and separate out along with the evaporation of alloy flux with the temperature increasing.

In the present work, the sample with superconductivity at 38.3 K was prepared in the Zn-Mg flux and magnetization curve above the zero line. The sample with superconductivity at 34.11 K was prepared in the Cd-Mg flux and magnetization curve below the zero line. $MgB_2$ prepared by these two methods was all flaky and irregular shape.

## 2. Experimental procedure

The starting materials of $MgB_2$ black powder with the weight of 0.5g and Zn-Mg (mass ratio 1:1, eutectic point ~623K), and Cd-Mg (mass ratio 1:1, eutectic point ~543K) alloy powder with the weight of 10g were well distributed and then enwrapped in the tantalum foil (thickness, 0.1mm) under the protection of argon atmosphere in the glove box. The tantalum package sealed in the silicon tube under gas protection and was put into the tube furnace.

The temperature program is shown in Figure 1. During the experiments, there were metal attached to the inner surface of the cold side quartz tube, indicating a more serious volatile flux. There is no difference in terms of quality between the final product and raw material $MgB_2$ initially placed in, which concluded that $MgB_2$ evaporation heat is very large, and also confirmed that this crystal growth method is feasible.

The sample was analyzed by X-ray powder diffraction (XRD) on an X-ray diffractometer (Rigaku D/Max-γ A) using Cu K-Alpha radiation (wavelength λ=1.54178 Å), superconducting quantum interference device (dc-SQUID) magnetometer using quantum design MPMS 5.5T, scanning electron microscope (SEM) using Hitachi S-2500.

## 3. Results and discussion

### 3.1 XRD characterization

The obtained crystals were analyzed by XRD, and the XRD pattern (Fig. 2) showed strong peaks of the $MgB_2$ crystals. There were a lot of $MgB_4$ among the Cd-Mg alloy flux (maximum temperature ~1270K) product, and a small amount of $MgB_4$ among the Zn-Mg alloy flux (maximum temperature ~1100K) product. This indicates that $MgB_2$ in the non-self-flux system, $MgB_2$ break down faster with the temperature rises as the following formula:

**$MgB_2 \rightarrow MgB_4 + Mg$**
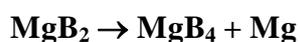

This is due to the low melting point alloy Zn-Mg or Cd-Mg flux to accelerate the evaporation of Mg in the decomposition products, so the higher the temperature, the more the $MgB_4$ content. Although melting point lower with the Zn-Mg or Cd-Mg alloy as a flux, little effect in helping to melt $MgB_2$. Low melting point metal can accelerate the flux evaporation compared with Mg-self-flux system [13], so it is necessary to require more and further experimental validation.

**3.2 SEM characterization**

The SEM image of the sample (Fig. 3) shows that the as-prepared $MgB_2$ crystals by these two methods exhibit a thin sheet, particle size of several microns in magnitude, it may be the result of defect of crystal growth system, or the fit technique is still not searched for because of being restricted of strict experimental conditions. However, these two metals should be the optional flux for $MgB_2$ crystal growth.

These crystals, which were detected by X-ray energy dispersion spectroscopic (EDS), as shown in Fig. 4, are composed of Mg, B (not present in the Fig.4 because of light element, including B, not being detected by EDS) and O. Although the reaction system (including crucible, tube furnace, starting materials, etc) is closely non-oxygen, the formation of MgO impurity phase is ineluctable [14]. According to the XRD and EDS analysis, there are three phase in products: $MgB_2$, $MgB_4$ and MgO.

**3.3 SQUID characterization**

The Meissner state magnetic susceptibility was measured under zero-field-cooled (ZFC) and field-cooled (FC) in an applied field of 10Oe, as shown in Fig.5. The as-prepared $MgB_2$ sample by using Cd-Mg flux showed an onset of superconductivity at about 34.11 K (Fig.5 (a)). Fig. 5a shows the FC signal to be about ~4% of the ZFC signal. Such a small FC signal reflects the very strong flux pinning character of single crystals. It is obviously that pinning force being from defects in the crystals counterworks movement of flux line. These defects form pinning centers. Therefore, the more the defects, the larger the pinning force on the vortices, the transition, which can be seen from the susceptibility curve. We rely on the current experimental means could not confirm that the defects exist in the crystal particles or powder because the particle size is small.

The $MgB_2$ sample by using Zn-Mg flux showed an onset of superconductivity at about 38.3 K (Fig. 5(b)). However, the susceptibility curve is above the zero line, this may be caused by impurities in the sample. Very little MgO impurities had no effect on the superconducting properties of $MgB_2$ [15]. If $MgB_4$ impurities impacted on $MgB_2$ superconducting properties, then the temperature susceptibility curve should be above the zero line in Fig.5a, but it did not happen. The $MgB_4$ low temperature magnetic properties were studied to clear the reasons.

### 3.4 $MgB_4$ low temperature magnetic properties

A small amount of $MgB_2$ powder was enwrapped in the tantalum foil under the protection of argon gas. The tantalum package was put into the tube furnace under the protection of argon atmosphere. According to the Mg-B binary phase diagram, the raw material was heated to ~1400K and stayed there for 1 h, and then cool to room temperature. The obtained sample was washed by dilute hydrochloric acid, and then washed several times with ethanol. The $MgB_4$ phase was identified by X-ray powder diffraction, as shown in Fig. 6, consistent with the standard diffraction card (JCPDS card 73-1014).

The low temperature susceptibility curve of $MgB_4$ was shown in Figure 7, and the curve above the zero line. Susceptibility reduced with increasing temperature, and there is no superconducting transition point.

MgB$_4$ was found in both samples, and MgB$_4$'s magnetization curve being above zero line was not superconductivity. No other component was observed by XRD and EDS. Up to now, the reason of magnetization curve difference between Zn-Mg flux and Cd-Mg flux is not made clear.

## 4. Conclusion

Using low melting point alloy flux method such as Zn-Mg and Cd-Mg as a new practice to grow MgB$_2$ crystal has been introduced. The sample with superconductivity at 38.3 K is prepared in the Zn-Mg flux and magnetization curve above the zero line. The sample with superconductivity at 34.11 K is prepared in the Cd-Mg flux and magnetization curve below the zero line. MgB$_4$ was found in both samples, and MgB$_4$'s magnetization curve being above zero line was not superconductivity. MgB$_2$ prepared by these two methods is all flaky and irregular shape; it may be the result of defect of crystal growth system, or the fit technique is still not searched for because of being restricted of strict experimental conditions. However, these two metals should be the optional flux for MgB$_2$ crystal growth. The study of single crystal growth is very helpful for the future applications.

## 5. Acknowledgments

This work was supported by the National Natural Science Foundation of China (No. 50802081, 51101133, 51101134) and Outstanding Young Scientist Award Found of Shandong Province (No. 2006BS04019).

Figure Captions

Figure 1 Temperature curve of $MgB_2$ crystals growth

Figure 2 XRD pattern of the as-prepared crystals of $MgB_2$ sample（a, Cd-Mg-flux；b, Zn-Mg-flux）

Figure 3 Image of the as-prepared $MgB_2$ sample: (a) Cd-Mg-flux; (b) Zn-Mg-flux

Figure 4 EDS of $MgB_2$ sample: (a) Cd-Mg-flux; (b) Zn-Mg-flux

Figure 5 Magnetic properties of the $MgB_2$ crystals sample: (a) Cd-Mg-flux; (b) Zn-Mg-flux

Figure 6 XRD pattern of the as-prepared crystals of $MgB_4$ sample

Figure 7 Low temperature susceptibility curve of $MgB_4$

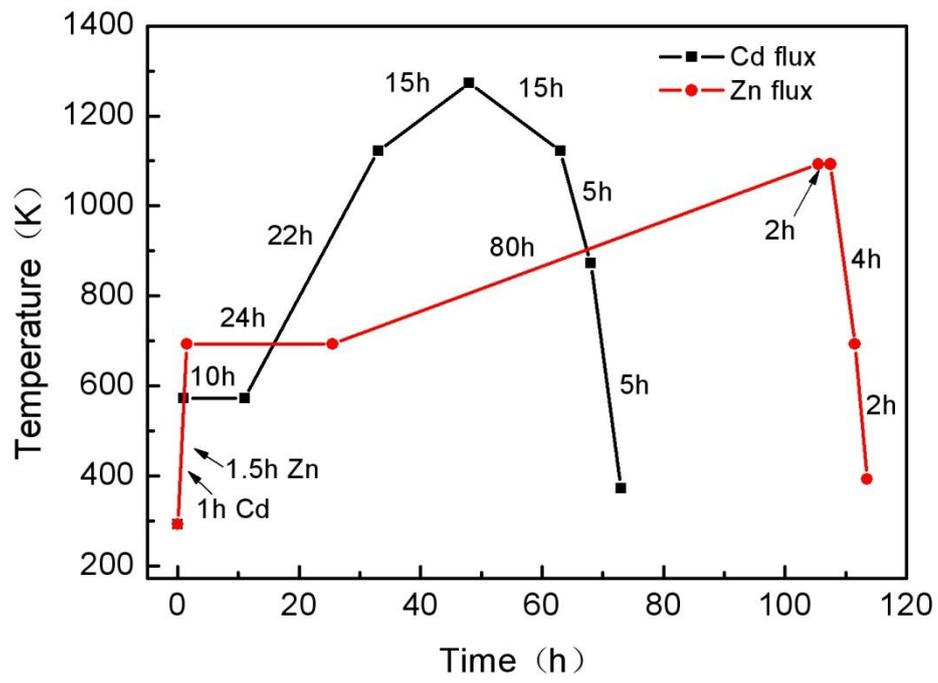

Figure 1 Temperature curve of $MgB_2$ crystals growth

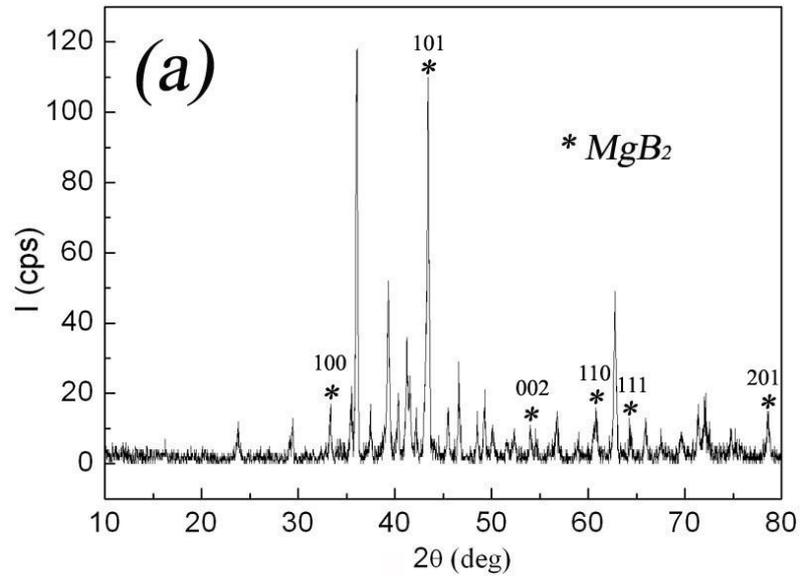

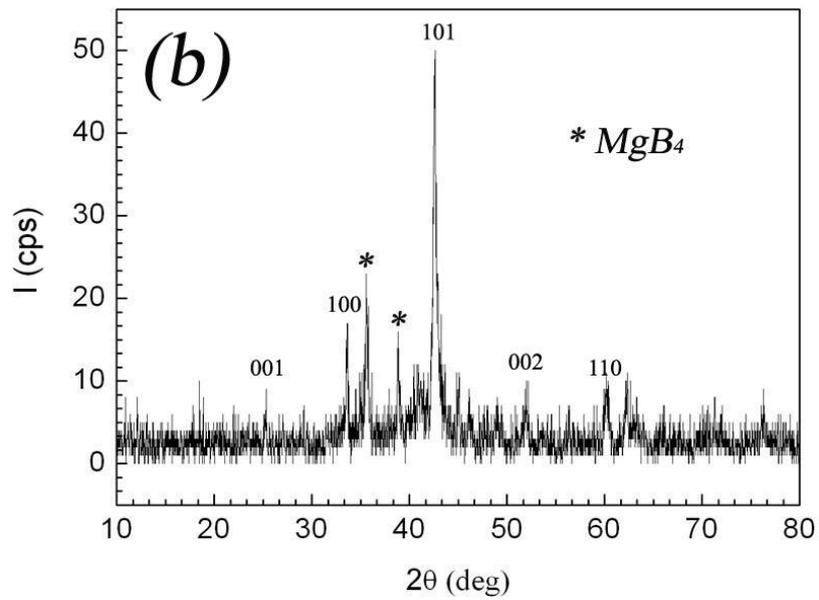

Figure 2 XRD pattern of the as-prepared crystals of $MgB_2$ sample（a, Cd-Mg-flux；b, Zn-Mg-flux）

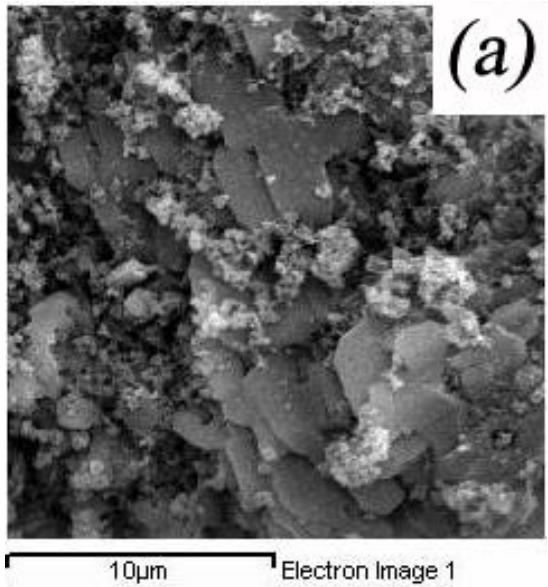

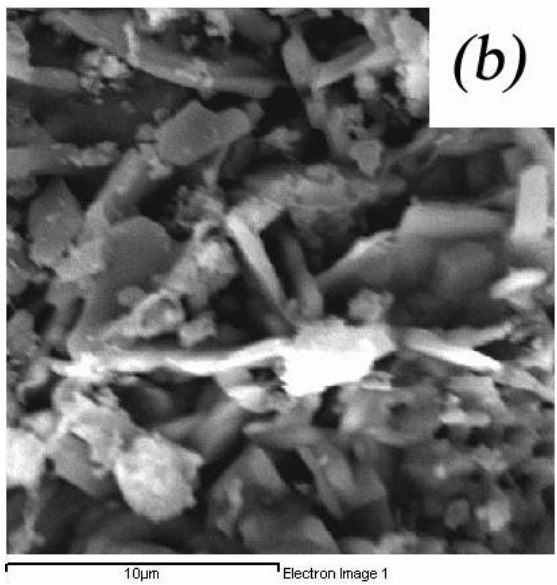

Figure 3 Image of the as-prepared $MgB_2$ sample: (a) Cd-Mg-flux; (b) Zn-Mg-flux

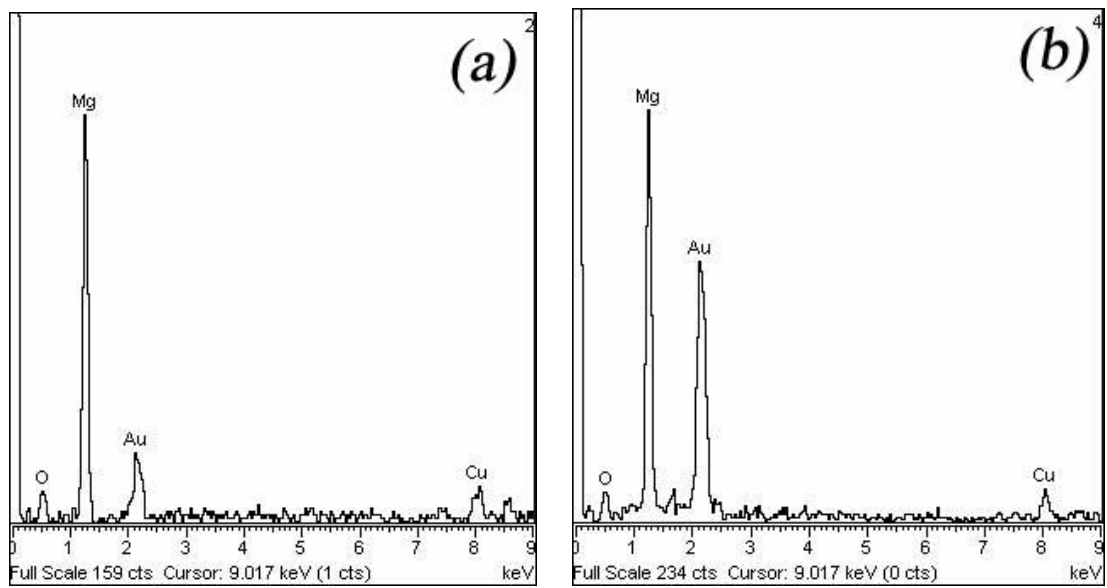

Figure 4 EDS of MgB$_2$ sample: (a) Cd-Mg-flux; (b) Zn-Mg-flux

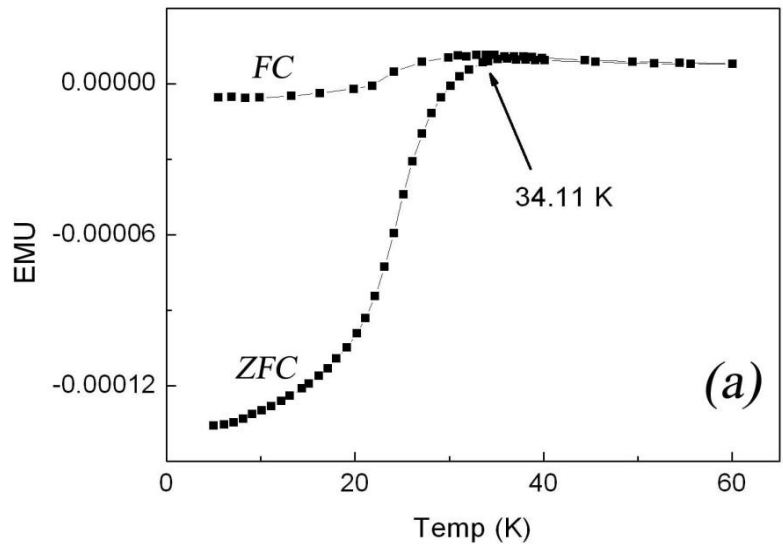

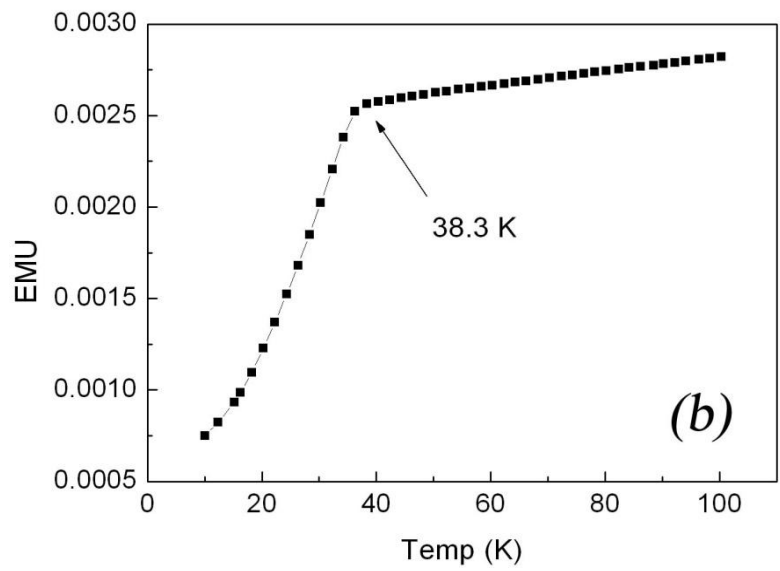

Figure 5 Magnetic properties of the $MgB_2$ crystals sample: (a) Cd-Mg-flux; (b) Zn-Mg-flux

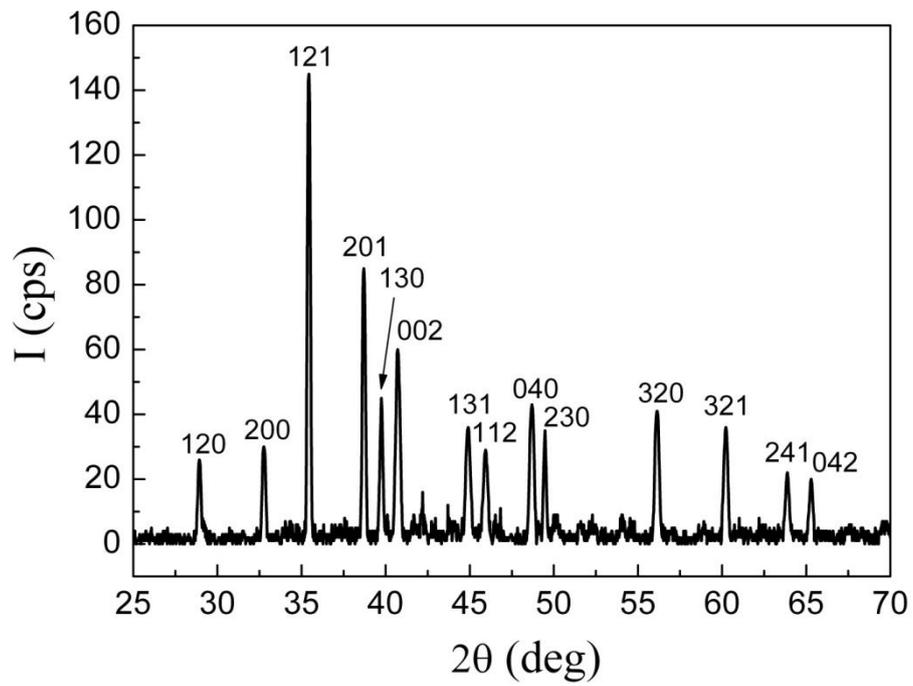

Figure 6 XRD pattern of the as-prepared crystals of MgB$_4$ sample

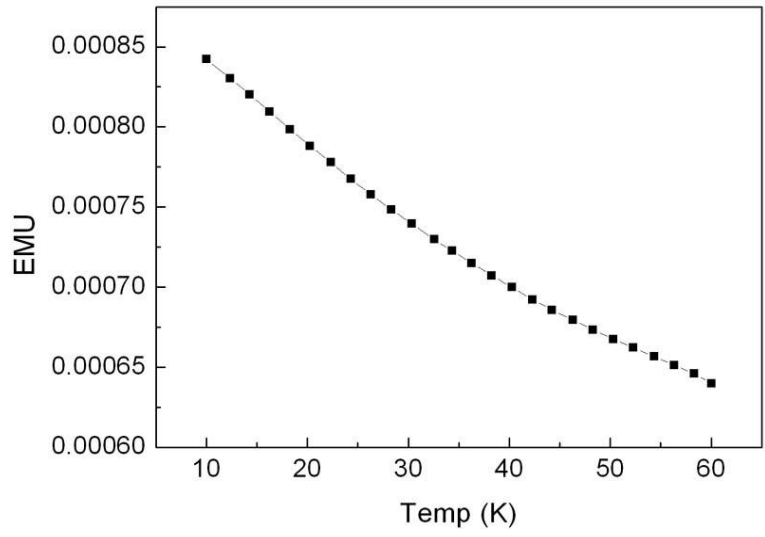

Figure 7 Low temperature susceptibility curve of $MgB_4$